# The Magic Word: A Coding Tutorial-Game to Engage Female Teenagers in App Design

**Bernadette Spieler,** *bernadette.spieler@uni-hildesheim.de*
Institute for Mathematics and Applied Informatics, University of Hildesheim, Hildesheim, Germany

**Naomi Pfaff,** *pfaff@student.tugraz.at*
Institute of Software Technology, Graz University of Technology, Graz, Austria

**Stefania Makrygiannaki,** *it164703@it.teithe.gr*
International Hellenic University, Nea Moudania, Greece

**Wolfgang Slany,** *slany@tugraz.at*
Institute of Software Technology, Graz University of Technology, Graz, Austria

## Abstract

Educational games are commonly used to motivate students and provide enhanced learning opportunities. Apps and mobile games play an increasingly important role in education and smartphones are part of the daily lives of most female teenagers: Half of mobile gamers are women and 64% of women prefer smartphones to other platforms. However, gender differences in playing behaviour and preferences raises concerns about potential gender inequalities when games are developed for education. In order to develop a tutorial game that suits the female target group and provides challenging tasks to solve, girls were involved at a very early stage of the development cycle and the idea was developed on the basis of surveys and focus group discussions. A first prototype of the game has been tested in a mixed-gender group to get feedback about the learning content, the worked examples, and the whole structure of the game. Finally, a tutorial game with six worked examples has been released in our Luna&Cat app, a programming tool that has been designed for our female target group in particular.

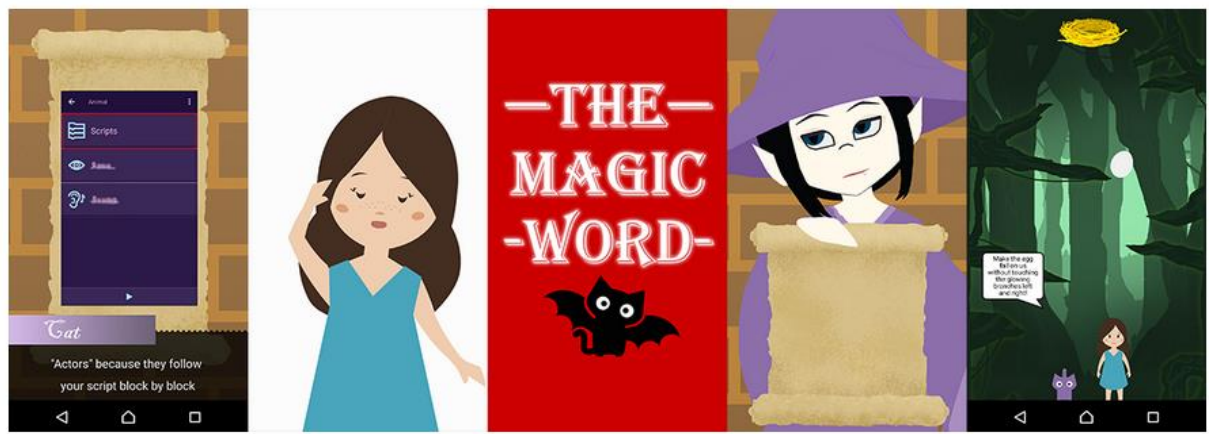

*Figure 1. Tutorial-Game for Luna&Cat: The Magic Word*

# Introduction

The acquisition of digital skills is more important than ever and represents a key professional qualification of future generations (European Commission, 2018). There is a great potential for young women to counteract the acute (and growing) shortage of qualified professionals in ICT (European Commission, 2016). However, the absence of female students who are interested in Information and Communication Technology (ICT) related fields can be observed at all levels of education as well as in the industry (European Statistics Eurostat, 2018). To allow women to improve their lives by offering them better career choices and more quality jobs, opportunities must be visible. Consequently, many girls leave school without any meaningful knowledge in Computer Science (CS), never quite understanding what CS is and how it relates to algorithmic thinking or problem-solving (Giannakos et al., 2014). These girls will be less likely to choose a career in CS or study it as a major. One accessible way is to catch their attention with attractive, supportive tools and games. To appeal to female teenagers in particular, a tailored version of our app Pocket Code, with the name "Luna&Cat" has been developed (Spieler and Slany, 2019). For this paper we design and develop a new coding tutorial for this app. The tutorial game should be extremely engaging and motivating for this challenging target group of younger female teenagers between 13 to 14 years old. At the same time, it should teach coding on small smartphone screens.

This paper is organised as follows: First, we provide an overview of the literature that informed the design of the tutorial game. The preferences of the game's target audience, the strengths of previous comparable games, and the psychological background of the game's learning mechanism are presented. In the following section, we outline the methods used to tailor the game to the game preferences of our target group as well as describing the results yielded by these methods. Next, we describe the first prototype and the procedure that was implemented in testing this prototype. The changes made to the finished version of the game based on our experiences with the prototype are also discussed in this section. Finally, our findings are presented and their implications and restrictions are discussed. Ongoing studies of the game are described and future research designs are suggested.

# Literature that Informed the Tutorial Game

For the development of the tutorial game, it was first essential to consider the preferences of our target group of female teenagers between 13 to 14 years old. Second, the game must be challenging and follow game design strategies and concepts, and third, it should deliver meaningful coding concepts and help girls to develop their own games in the future. Beside explaining important game design concepts and gender conscious strategies in games, this section provides an overview to Constructionism and tutorial games, and highlights challenges in problem- based learning by considering cognitive load and applying worked examples.

## Gaming Perceptions and Experiences in Teenager Girls

The ESA annual report contains statistics on video game player demographics (ESA, 2006-2013). Here, we see a slight change from 2006, where female players accounted for 38% of video game players, to a total of 45% female video game players in 2013. The flow state, described by Csikszentmihaliy (1975), is the state in which people are fully absorbed in their task, forget space and time, and only care about the activity - they are in a flow experience. This state of mind is very common while playing games. When comparing self-documented data with actual data from game servers, it was found that women underestimated their playing time by an average of 3 hours per week, while men overestimated their playing time by an average of 1 hour per week (Williams et al., 2009a). That users reach this state in the context of an educational tutorial game is very desirable for developers. Furthermore, William, et. al. (2009b) find out upon examination that men were significantly more motivated by achievement and women were significantly more motivated by social and immersion factors such as geographic exploration, role-playing, avatar customization, and escapism. In describing game style preference, Kinzie and Joseph (2008) suggest that girls at a mean age of 12 prefer creative and explorative aspects of a game, and Quaiser-Pohl, Geiser and Lehmann, (2006) continue that female high school girls choose more non-players or logic/skill training games.

To conclude, a closer look at game designs in reference to the different motivations for gameplay discussed above, as well as full consideration of the different gaming style preferences between gender, can prevent developers from excluding potential player groups.

## Constructionism and Tutorial-Games

Tutorial games are available for many computer games, but rarely for learning programming. The few examples available are usually playable for a few minutes and not very exciting. A great exception to this common problem that appeals to teenagers is Nintendo's tutorial game in "Wario Ware[42]: Do It Yourself." Another example tutorial game for coding designed especially for girls is the one with Wonder Woman[43].

The constructionist theory by Semour Papert (1985) explains the benefit if students use computational technologies to construct knowledge through the act of constructing personally meaningful projects. In Mindstorms (1985), Papert suggests the use of MicroWorlds in which learning about specific principles occur. Tutorials can be organized in such a way that instructions help students to develop technological knowledge to further acquire the skills to express themselves and their ideas through new tools (Stager, 2001). The Luna&Cat tutorial will enable girls to feel intellectually powerful by solving challenges after challenges, perhaps for the first time in their lives.

## Problem Solving Instruction

**Problem Schemas** are mental representations of solutions to a type of problem. They contain the operators that can be applied to a problem, the problem states that lead from the initial problem to its solution, and the effect of operators on the problem during different problem states (Sweller and Cooper, 1985). This information can be used to solve any problem belonging to the same category. Expertise in a problem domain can be understood as the number of problem schemas acquired by an individual. The main learning outcome of problem solving is the acquisition of problem schemas.

**Cognitive Load:** To solve a problem, one has to mentally represent all components of the problem and manipulate them by applying operators to the problem (van Gog et al., 2006). The strain a problem puts on cognitive resources is the cognitive load imposed by this problem. The cognitive load imposed by problem solving affects novices and experts differently. Experts have automatized the mental representation of the problems' components so that they are left with sufficient cognitive resources to acquire problem schemas during problem solving. Novices' representation of the problems components is not automatic and therefore imposes a high cognitive load on their working memory.

**Worked Examples** are solved problems. For example, in Algebra, a worked example is a line by line documentation of every problem state from the initial problem to its resolution. They show learners how the initial problem state is transformed to the goal state through the use of operators. By definition, worked examples contain all the information that is necessary to form a problem schema. Worked examples impose a lower cognitive load than problem solving while providing the same information (van Gog et al., 2006). Therefore, learners have more cognitive resources available for the processing of problem schemas, leading to a better retention of problem schemas. This has been shown to increase learning outcomes in both expert and novice learners when compared to problem solving (Carroll, 1994; Sweller and Cooper, 1985; van Gog, Kester and Paas, 2011; van Gog et al., 2006). Novice learners receive an even greater benefit from the study of worked examples than experts because the high cognitive load that is imposed by problem solving impacts their ability to acquire problem schemas acquisition more severely (Sweller and Cooper, 1985). Therefore, worked examples are especially beneficial to our target group.

---

[42] Wario War: https://www.youtube.com/watch?v=4lSQqUoqQ9Q&feature=youtu.be&t=88
[43] Wonder Women: https://www.madewithcode.com/projects/wonderwoman

## Research Design

For the design and story of the tutorial game a bottom up approach was used by developing personas first. A persona is a description of user characteristics and her or his aims (Cooper, 2003). According to Cooper, a persona should be presented in text and/or image and it is usually generated to help designers to understand, describe, and define user preferences and behavior patterns. The data for the personas for this tutorial game has been constructed on the basis of questionnaires and focus discussions. In May 2018, a survey was created which included questions surrounding girls' game preferences, internet use (mobile games, social media), and general information (hobbies, interests, TV series, movies). The survey was handed out to 21 female students in Grade 4 (between 13 to 14 years old). In addition, a focus group discussion in which four students attended was conducted in order to ask more specific questions about the topics which emerged.

Furthermore, in June 2019, a first prototype of the tutorial game which already included three worked examples, was tested in a class of 24 students. This was important to get insights on how a tutorial game could be integrated in our Luna&Cat app as well as how to approach further difficulties during solving the tutorial.

## Building a Gender-Conscious Tutorial Game by using Personas

The app Pocket Code has been developed at the Graz University of Technology in Austria under the Catrobat association (https://catrobat.org). It has a media-rich programming environment for teenagers to learn coding with a visual programming language very similar to the Scratch (https://scratch.mit.edu/) environment, which has been used for creating games and apps (Slany, 2014). Pocket Code is freely available on the Google Play Store (https://catrob.at/pc) and on iTunes (https.//catrob.at/PCios) and allows for the creation of games, stories, and many types of other apps directly on phones, thereby teaching fundamental programming skills. In addition, a new version of our app with the name Luna&Cat is available on Google Play (https://catrob.at/luna) since April 2019. This version has been developed by considering gender-sensitive aspects from gender studies with the goal to reinforce female teenagers. To engage more female teenagers in coding, a tutorial game has been created especially for this app version. The focus was on the conception, design, and development and testing the tutorial game.

Personas were used to determine the game genre of the tutorial and the narrative genre of its storyline (Cooper, 2004). In an interaction with a digital game, the primary goal of any user is to have fun. Students were categorised based on their requirements for fun; the game genres that students named as their favorites were used to for categorisation. Personas were developed from the answers given by group members on our questionnaire. We then constructed a questionnaire filled out by the persona by inserting real answers of group members belonging to the same group into a single questionnaire.

The first group was used to create the primary persona. This group favours "Jump&Run" games. Their favorite feature of their favorite games is that "you have to be clever and lucky". They prefer light games that emphasise manual and perceptual skill over story and strategy. Several members of this group listed "Candy Crush" as their favorite game and two students stated that they liked "Candy Crush" because "you have to think". The persona for the "Jump&Run" group is **Natalie**. Her favorite movies are the Harry Potter movies. Her favorite singer is Billie Eilish and she has written Billie Eilish's darker song lyrics into her the sidelines of her school notebooks in dramatic calligraphy. She likes the tone and imagery of "Alice in Wonderland" for the same reasons she likes Billie Eilish. She is very good at braiding hair. She has calm hands and is good at fiddling with bits, which is why her family asks her to open small clasps on necklaces or repair small earrings.

The second group was used to create the secondary persona. Its members stated that their favorite game genre was adventure games. Its quote regarding its favorite feature of its favorite game was "The world is huge, so you can always discover something new. It's not boring because you can make your own free decisions". This group did not like games which had issues such as "not enough space, not free enough, no more quests". The persona for the adventure game group is **Laura**. Laura's favorite book is the Alexander Rider series. She enjoys physics and the feeling of understanding a new concept. She likes Marvel movies and StarWars. She also associates the science fiction elements of StarWars and Marvel with physics and the sciences and therefore feels a connection to their audience. She is also interested in and fascinated by the science fiction technology featured in these movies.

The design of the game combined the interests of these personas. It is true that it is considered good practice to keep personas as specific as possible as it is thought to be better to create a product that is very appealing to a small group than to create a product that is mildly appealing to a large group (Cooper, 2004). However, our study was addressing the preferences of a previously specified group, that of 13 to 14 year old girls, so that it was possible to combine personas without significantly broadening the target group.

The exercises in the first game would appeal to Natalie. They are puzzles and their increasing difficulty gives Natalie a sense of self-efficacy. Programming in visual languages such as the Catrobat language involves perceptual skills such as the recognition and memorisation of patterns. The individual tasks designed for her were to serve as minigames which are lighter and closer to games like "Candy Crush". These minigames were placed within a larger and more immersive adventure game.

The storyline of Luna's quest was designed for Laura, but it would appeal to both Laura and Natalie. Laura would enjoy it because of its immersive effect and the fictional world created by the narrative; both of these features are central to adventure games. Laura would prefer if there was more choice of actions within the game; this should be considered in future versions of the game. Laura would be excited to learn to code because of the connection between programming and the development of new technologies. She would be intrinsically motivated to understand the worked examples.

## Case Study: Application of the Tutorial Game

The prototype focused on having users test some in-app features on an educational application, while on the other hand, the finished version concentrated on having users code missing parts of the story in a game. A first user test of the prototype involved 24 students (11 female, 13 male) in an academic high school in Graz, Austria. The goal was to test if the concepts delivered by the tutorial could be transferred to similar tasks. Therefore, half of the class was randomly assigned a tutorial game that included animations and pictures of code snippet to explain the worked example and the other group had only a description in the form of a text, see Figure 2. The results of this first test did not produce a large range in values, making the interpretation of the effect of the two conditions difficult to decipher. The mean number of errors made by the text example group was 0.83 while the mean number of errors made by the worked example group was 1.60. The text example group took an average of 9 minutes to solve the problems while the worked example group took 11 minutes. One possible interpretation of these results would be that the worked example group produced more errors because they wrote more code. This is supported by observations during the grading of the programs written by students. Students in the text example group did not finish their programs or handed in empty programs. As the method of grading had previously been defined as counting errors it did not allow for this to be included in grading.

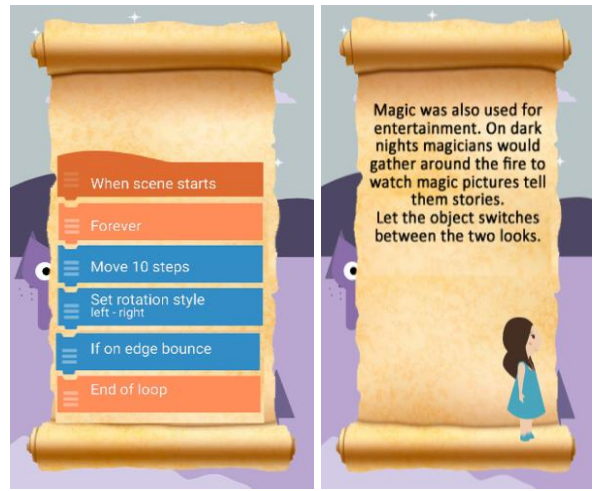

*Figure 2: Left: animations and code; Right: text description only.*

## The Magic Word: A First Concept

The final version of the game displays a major and drastic improvement in comparison with the prototype. The storyline and teaching method were kept the same theoretically, as friendlier graphics and easier to use interface was developed. General add-ons to the game were a menu for easier navigation through the game, see Figure 3.

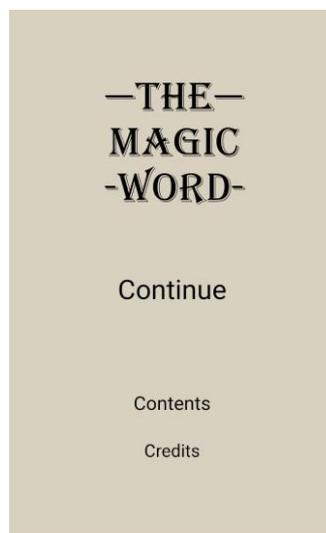

*Figure 3. The Magic Word: Main Menu*

The story of the game blends programming into the game world and makes programming ability central to the success of the main character's quest. The story is also designed to get the user emotionally invested in the main character's success. In the prologue the user helps Luna's mentor Cat to steal a book. This book appears in Lunas room. We see Luna examine to book. She is interrupted by Noodles, her puppy. They play and during the game, the puppy falls through the book which is a portal to a magical world. Cat appears and explains that Luna is a descendant of magicians from this world and that it is time for her to learn magic. The magic Cat teaches Luna is directed by the user through coding. The "spells" Cat shows Luna are worked examples of code. The user's ability to apply the code to the problem in the game world determines the success of the spell cast by Luna. As the story progresses Luna and Cat get closer and closer to Noodles. The obstacles in their way can only be removed by magic. Table 1 presents the worked examples:

**Table 3: Worked examples (WE) of the tutorial game**

| 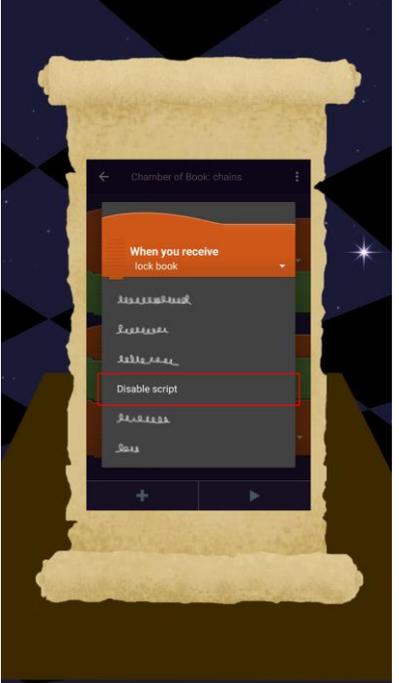 | 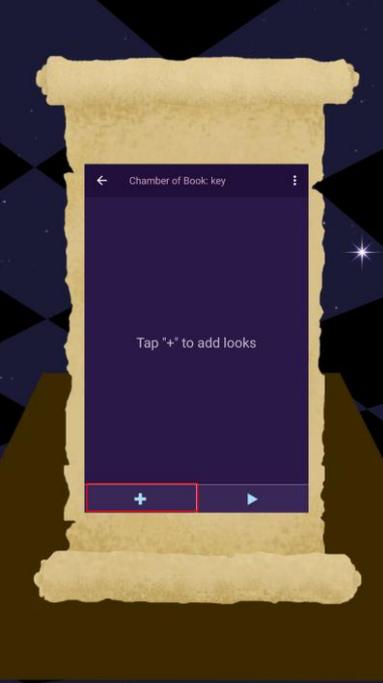 | 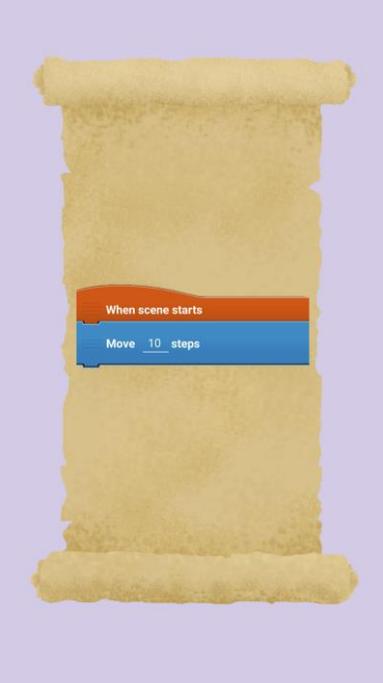 |
|---|---|---|
| WE 1: disable code blocks. | WE 2: add looks to game objects. | WE 3: function of motion blocks. |
| 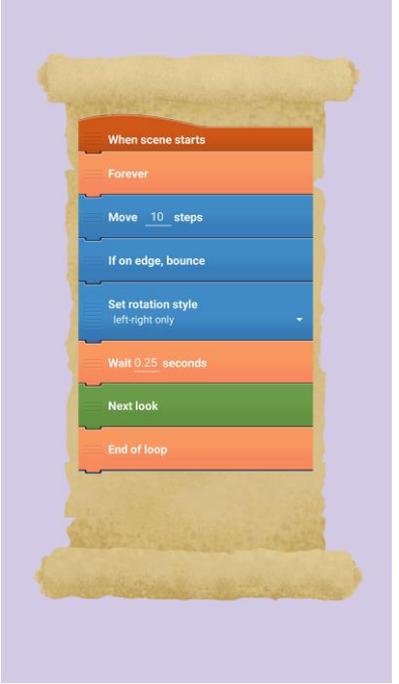 | 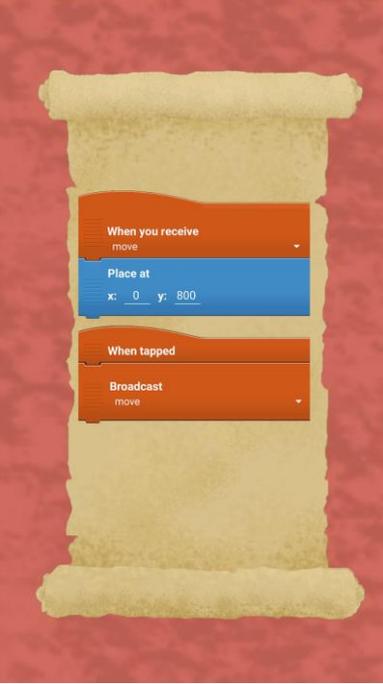 | 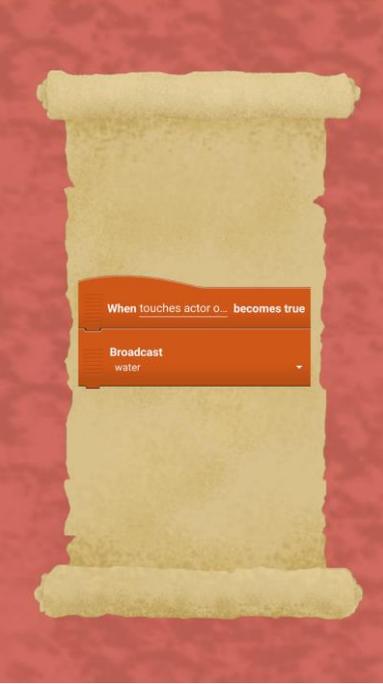 |
| WE 4: function of loops. | WE 5: function of broadcasts. | WE 6: function of conditions. |

| 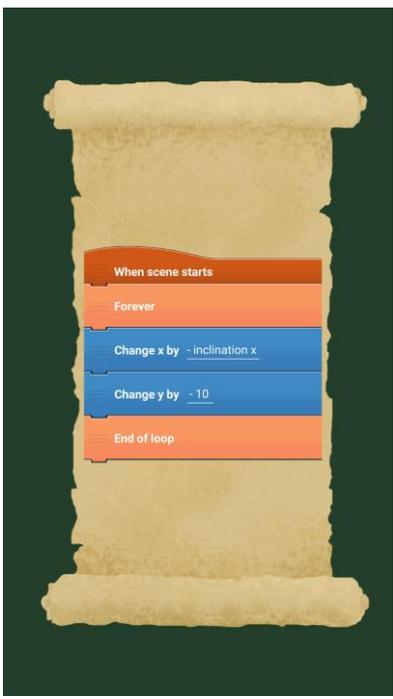 | 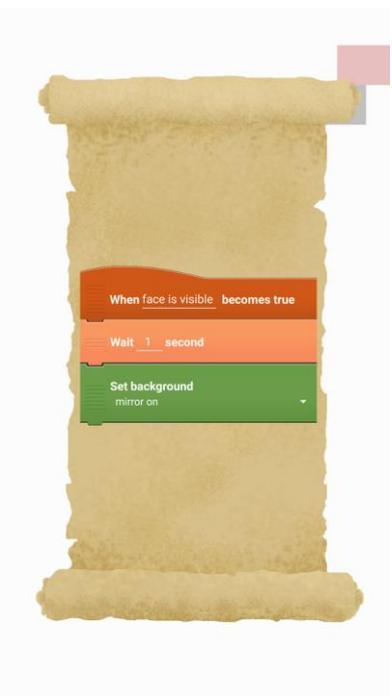 | 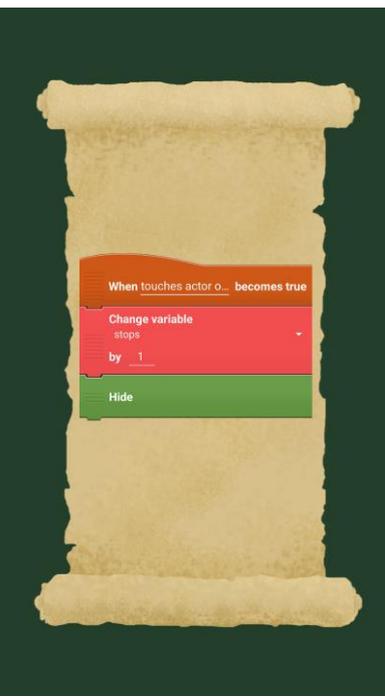 |
|---|---|---|
| WE 7: function of motion sensors. | WE 8: function of visual sensors. | WE 9: function of variables. |

## Discussion & Conclusion

At the time of writing, a study of the tutorial's effect on programming ability is still in progress. This one following the first case study, tested more students and counted the number of correct code blocks in order to account for this issue. For the first case study the usability issues with the prototype definitely had a negative impact on learning outcomes. The clear step-by-step instructions provided in the finished version of the game have resolved this issue. There was also confusion about the object-oriented design of the Luna&Cat app. In the finished version this issue was addressed through a unit that is dedicated to explaining this concept as well as other important functionalities of the app. The effect of the tutorial's learning mechanism should be more clearly visible in future studies which will use the improved final version of the tutorial.

Testing the game with a larger sample would provide a clearer picture of the tutorial's learning outcome. While the tutorial's learning outcome is central to an assessment of its quality, it is true that the tutorial was only intended to teach basic programming principles that could then be built on through the active use of the Luna&Cat app. While the understanding of basic programming principles is a prerequisite for the successful use of the the Luna&Cat app, willingness to engage with the app is also needed. Future studies should investigate the tutorial's effect on participants' willingness to engage with programming and the Luna&Cat app. For example, the tutorial's motivational effects could be investigated by testing the usefulness of the tutorial game as a predictor of future programming experience in a longitudinal study.

BibTex entry:

```
@InProceedings{SPIELER2020,
author = {Spieler, B., Pfaff, N., Makrygiannaki, S., and Slany, W.},
title = {The Magic Word: A Coding Tutorial-Game to Engage Female Teenagers in App Design},
conferencetitle = {Proceedings of the 2020 Constructionism Conference},
isbn = {978-1-911566-09-0},
location = {Dublin, Ireland},
month = {26-29 May 2020},
year = {2020},
pages = {556 - 564}
}
```